\documentclass[prb,twocolumn,showpacs,superscriptaddress]{revtex4-1}

\usepackage{graphicx,amssymb,amsmath,color,bbm,psfrag,bm}
\usepackage[utf8]{inputenc}
\usepackage{color}
\usepackage[normalem]{ulem}

\begin{document}
\definecolor{darkgreen}{rgb}{0,0.5,0}

\newcommand{\jav}[1]{\textcolor{red}{#1}}

\title{Nuclear spin-lattice relaxation time in TaP and the Knight shift of  Weyl semimetals}
\author{Zolt\'an Okv\'atovity}
\affiliation{Department of Theoretical Physics and MTA-BME 
 Lend\"ulet Topology and Correlation Research Group, Budapest University of Technology and Economics, Budapest, Hungary}

\author{Hiroshi Yasuoka}
\affiliation{Max Planck Institute for Chemical Physics of Solids, 01187 Dresden, Germany}
\author{Michael Baenitz}
\affiliation{Max Planck Institute for Chemical Physics of Solids, 01187 Dresden, Germany}

\author{Ferenc Simon}
\affiliation{Department of Physics and MTA-BME Lend\"{u}let Spintronics Research Group (PROSPIN),  Budapest University of Technology and Economics}

\author{Bal\' azs D\' ora}

\email{dora@eik.bme.hu}

\affiliation{Department of Theoretical Physics and MTA-BME 
 Lend\"ulet Topology and Correlation Research Group, Budapest University of Technology and Economics, Budapest, Hungary}

\date{\today}

\begin{abstract}
We first analyze the recent experimental data on the nuclear spin-lattice relaxation rate of the Weyl semimetal TaP. We argue that its non-monotonic temperature dependence is explained by the temperature dependent chemical potential of Weyl fermions.
We also develop the theory of the  Knight shift in Weyl semimetals, which contains two counteracting terms. The diamagnetic term follows $-\ln[W/\max(|\mu|,k_BT)]$ with 
$W$, $\mu$ and $T$ being the high energy cutoff, chemical potential and temperature, respectively, and is always negative. 
The paramagnetic term scales with $\mu$ and changes sign depending on the doping level.  Altogether, the Knight
shift is predicted to vanish or even change sign upon changing the doping or the temperature, making it a sensitive tool to identify Weyl points.
We also calculate the Korringa relation for Weyl semimetals which shows an unusual energy dependence rather than being constant as expected for a non-interacting Fermi system.
\end{abstract}

\maketitle

\section{Introduction}

With the advent of topological insulators, the observation of many fascinating phenomena became possible\cite{hasankane,zhangrmp}, including
the magnetoelectric effect, axion electrodynamics, Majorana fermions.
 In their bulk, these materials resemble to a normal insulator, but their surfaces or edges
host metallic states, which are protected by the underlying topology. In this respect, they are regarded as the descendant of quantum Hall states, which is manifested in e.g. the quantized spin-Hall conductivity in spin-Hall insulators\cite{konig}.

The above story can further be twisted by designing materials whose bulk metallicity is protected by topology. A topological metal
in 3D is incarnated in  Weyl semimetals\cite{herring,wan,murakami,BurkovPRL2011}. 
The protection of metallic behaviour is best visualized in momentum space, where 
a Weyl point may be regarded as a magnetic monopole \cite{turner}. These objects appear pairwise, and can only be annihilated by
colliding two monopoles with opposite topological charge into each other.
Due to the non-trivial topology, Weyl semimetals also feature a variety of extraordinary phenomena such as the {chiral anomaly or} the anomalous Hall conductivity \cite{turner, burkov}.

\begin{figure}[h!]
\includegraphics[width=8cm]{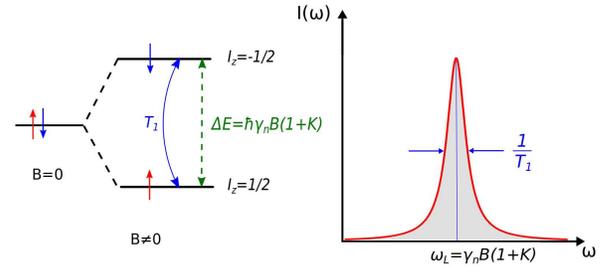}
\vspace*{-4mm}
\caption{Sketch of NMR: nuclear spin states are split by an external magnetic field $B$, whose energy scale  is measured together with the relaxation time.}
\label{nmrbasic}
\end{figure}

While surface sensitive probes such as STM or ARPES capture the physics of protected surface states, i.e. Fermi arcs for Weyl semimetals\cite{fermiarc1,fermiarc2},
bulk probes also provide valuable information about the electronic structure. Among these, 
 nuclear magnetic resonance (NMR) technique has long been known\cite{winter,abragam,SlichterBook} to reveal a plethora of information about the
 electronic or other degrees of freedom, through which nuclear spins relax. 
For example, the exponential vs. power law temperature dependence of the relaxation time, $T_1$ (see Fig. \ref{nmrbasic}) in a superconductor contains information about the 
structure of the superconducting gap and its possible nodal structure,
while the position of the resonance, i.e. the Knight shift $K$, depicted in Fig. \ref{nmrbasic}, distinguishes between 
singlet and triplet pairing\cite{HebelSlichter,maeno}. In materials whose superconductivity is mediated by spin-singlet pairing, the Knight shift drops with decreasing 
temperature, while it stays at its normal state value for spin-triplet Cooper pairs.

At  the heart of the NMR lies the hyperfine coupling, describing the interaction between nuclear spins and the surrounding medium.
In Ref. \onlinecite{okvatovity}, we determined the hyperfine interaction for Weyl semimetals using an "ab-initio" treatment of the 
low energy effective Hamiltonian. This allowed to show that the spin-lattice relaxation rate
is anomalous in Weyl semimetals and does not follow the behaviour expected from the density of states.
Instead of a $1/T_1T\sim E^4$ scaling with $E$ being the maximum of temperature ($k_BT$) and chemical potential,
the  nuclear spin relaxation rate scales in a graphene like manner\cite{doranmr1} as $1/T_1T\sim E^2\ln(E/\omega_0)$ with $\omega_0$ the nuclear Larmor frequency.
{In Sec. II, we introduce the model developed in Ref. \onlinecite{okvatovity} to set the stage for the subsequent analysis.
In Sec. III, we first recapitulate our previous work on the nuclear spin relaxation time, and then apply the result to the recent nuclear quadrupole relaxation data on 
TaP\cite{yasuoka} and demonstrate that by taking the temperature dependence of the chemical potential into account, we are able to describe the salient features of the
experimental data.
In Sec. IV, we provide a similar ab-initio evaluation of  the Knight shift in Weyl semimetals as well, }
which reveals rich behaviour depending on the conspiracy of the chemical potential and temperature. Namely, it can cross over between diamagnetic and paramagnetic behaviour 
by tuning them, respectively.
The Korringa relation of a Fermi liquid, {studied in Sec. V}, is not satisfied due to the strong spin orbit coupling, which is essential to induce Weyl points.

\section{Hyperfine interaction in Weyl semimetals}

Following Ref. \onlinecite{okvatovity}, we rederive the hyperfine interaction in Weyl semimetals.
By focusing on the low energy excitations, the Hamiltonian of Weyl semimetals is written as
\begin{equation}
H=v_{F}(p_x\sigma_x+p_y\sigma_y+p_z\sigma_z),
\label{hamilton}
\end{equation}
{Here, the physical spin of the electron is represented by the Pauli matrices ($\sigma$'s), and $v_{F}$ is their Fermi velocity}, typically\cite{neupane,chiral2} 
of the order of $10^{5}-10^{6}$~m/s.
Its dispersion relation is also linear in momentum, as is usual for zero mass Weyl fermions in arbitrary dimension (e.g. for graphene as well\cite{rmpguinea}) as
\begin{equation}
\varepsilon_\lambda({\bf k}) = \lambda v_{F}\hbar\vert{\bf k}\vert
\label{weylenergy}
\end{equation}
with $\lambda=\pm$ and $k=|\bf{k}|$ for the length of the 3D momentum.
The spinor eigenfunctions are written as
\begin{subequations}
\begin{gather}
|{{\bf k},+}\rangle=\begin{bmatrix} \cos{\left(\frac{\vartheta_{\bf k}}{2}\right)} \\\sin{\left(\frac{\vartheta_{\bf k}}{2}\right)}
\exp(i\varphi_{\bf k}) \end{bmatrix}
\label{weylfunction1}\\
|{{\bf k},-}\rangle=\begin{bmatrix}\sin{\left(\frac{\vartheta_{\bf k}}{2}\right)} \\-\cos{\left(\frac{\vartheta_{\bf k}}{2}\right)}\exp(i\varphi_{\bf k}) \end{bmatrix}.
\label{weylfunction2}
\end{gather}
\label{weylfunction}
\end{subequations}
{The $+$ and $-$ components in Eqs. \eqref{weylfunction}  correspond to positive and negative eigenenergies, respectively, and 
$\varphi_{\bf k}$ is the azimuthal angle in the ($k_x$,$k_y$) plane and $\vartheta_{\bf k}$ is the polar angle made from the $k_z$ axis in a spherical coordinate system.}

In Ref. \onlinecite{okvatovity}, the standard route outlined in Refs. \onlinecite{abragam,doranmr1} 
was followed to obtain the hyperfine interaction. After representing the nuclear spin as a dipole with dipole moment ${\bf m}=\hbar\gamma_{n}{\bf I}$, its vector potential is
\begin{gather}
{\bf A} = \frac{\mu_0}{4\pi}\frac{{\bf m}\times {\bf r}}{r^3}=\frac{\mu_0}{4\pi}\hbar\gamma_{n}\frac{{\bf I}\times {\bf r}}{r^3}.
\label{vectorpot}
\end{gather}
{Here $\gamma_{n}$ is the nuclear gyromagnetic ratio and   $\mu_0$ is the vacuum permeability. The vector potential, stemming from the dipole,
appears  in the Hamiltonian through the Peierls substitution as ${\bf p} \rightarrow {\bf p}+e \bf A$ with $e>0$ the elementary charge, }
and its magnetic field, $\nabla\times \bf  A$ through the Zeeman term.

Using this "ab-initio" treatment of the nuclear spin within the low energy effective Hamiltonian of Eq. \eqref{hamilton},  
the hyperfine interaction between a localized nucleus and the surrounding Weyl fermions after some lengthy calculation\cite{okvatovity}
reads  as\footnote{{Some misprints are corrected compared to Ref. \cite{okvatovity}.}}
\begin{gather}
 {H}_{HFI}=\frac{\mu_0}{q^2}\gamma_{n}\hbar{\bf I}\cdot\left[
iev_{F}\left({\bf q}\times\boldsymbol{\sigma}\right)
-\frac{g\mu_B}{2}\left({\bf q}\times \left({\bf q}\times \boldsymbol{\sigma}\right)\right)\right],
\label{hfift}
\end{gather}
where the momentum transfer between the incoming ($\bf k$) and outgoing ($\bf k'$) electron, which gets scattered off the localized spin, is ${\bf q}={\bf k}-{\bf k'}$. 
The first and second term are the orbital and the spin part of the hyperfine interaction.
The first one is the Fourier transform of $\bm\sigma\cdot\bf A$, and its $\bf q$ dependence comes from the Fourier transform of Eq. \eqref{vectorpot}, as shown in Ref. \onlinecite{okvatovity}.
The second term is the Fourier transform of the magnetic field from Eq. \eqref{vectorpot}, $\bf B=\nabla\times A$, which explains the extra $\bf q\times$ factor
compared to the first term.
The peculiar feature in Eq. \eqref{hfift} is the $ev_{F}/q$ divergence of the orbital hyperfine coupling for $q\rightarrow 0$.
The second term containing $g\mu_{B}$ remains finite in the small $q$ limit, since both the numerator and the denominator vanish with $q^2$.

The above Hamiltonian neglects structures on an atomic length scale, and is the universal contribution from Weyl fermions, 
valid in the low energy long wavelength limit. Additional short range terms to the hyperfine
coupling can also arise from short range processes within the real space unit cell\cite{fischer,lunde}, 
which can be taken into account by considering the lattice periodic Bloch
wavefunction as well. This contribution is, however, non-universal and depends on the actual geometry of the lattice and the real space unit cell, which hosts Weyl fermions.
Nevertheless, 
the lattice periodic Bloch wavefunction, $u_{\bf k}({\bf r})$ can be Fourier expanded in terms of reciprocal lattice vectors, $\bf G$ as $u_{\bf k}({\bf r})=\sum_{\bf G}c_{\bf k}({\bf G})e^{i{\bf Gr}}$, and the Fourier transform yielding Eq. \eqref{hfift}
would now contain ${\bf q}+\Delta{\bf G}$ instead of $\bf q$, and $ \Delta \bf G$ is the reciprocal lattice vector difference of  two Bloch states.
 However, the $\Delta {\bf G}=0$ contribution is  present in general and gives the most dominant contribution in the small 
$\bf q$ limit, as we detail it in Appendix A.  Therefore, we focus only on this as the universal signature of Weyl fermions, and neglect the non-universal structure on atomic length scale.
Since many different lattices with distinct unit cells give rise to Weyl fermions, it is important to focus on the universal long wavelength contribution
without the non-universal  short range pieces.
The same approach was found to describe the NMR relaxation rate and Knight shift on graphene\cite{doranmr1} and 
as we show below, this  accounts successfully for the spin relaxation rate in TaP Weyl semimetal.

\section{Nuclear spin relaxation in the Weyl semimetal TaP}

In Ref. \onlinecite{okvatovity}, 
we derived the spin-lattice relaxation rate of Weyl-fermions from  an effective low energy description of the fermionic excitations. 
Surprisingly, the dominant contribution at low $T$ and $\mu$ comes from the orbital
part of the hyperfine interaction, which usually gives a small contribution in normal metals.

The relaxation time {was evaluated as\cite{okvatovity}}
\begin{gather}
\frac{1}{T_1}=\frac{\pi\mu_0^2\gamma_{n}^2}{4 v_{F}(2\pi)^6}\int\limits_{-\infty}^{\infty}dk  
\dfrac{\left(kev_{F}\right)^2 F\left({|k|}/{k_0}\right)}{\cosh^{2}\left[(\hbar v_{F} k-\mu)/{2k_B T}\right]}
\label{weylrelax},
\end{gather}
where  $k_0={\omega_0}/{v_{F}}$ is the Larmor wavenumber, 
 $\omega_0=B\gamma_{n}$ is the nuclear Larmor frequency,  which is the smallest energy scale of the problem
due to the heavy mass of the nucleus,
$\gamma_{n}$ is the gyromagnetic ratio of the studied nucleus and
 the dimensionless functions $F(x\rightarrow 0)\approx 52.7 \ln\left(2{x}\right)$. From Ref. \onlinecite{maebashi}, the numerical constant $52.7$ is $(4\pi)^2/3$.

By performing the remaining integral, we eventually obtain
\begin{gather}
\frac{\hbar}{T_1k_BT}=\frac{52.7\pi\mu_0^2\gamma_{n}^2 e^2}{(2\pi)^6 v_{F}^2}\times\nonumber\\
\times\left\{
\begin{array}{cc}
\left(\dfrac{k_B T}{\hbar}\right)^2 \dfrac{\pi^2}{3}\ln\left(\dfrac{4 k_B T}{\hbar\omega_0}\right),& \textmd{ } \mu\ll k_BT \\
\left(\dfrac{\mu}{\hbar}\right)^2\ln\left(\dfrac{2 \mu}{\hbar\omega_0}\right),& \textmd{ } \mu\gg k_BT.
\end{array}\right.
\label{t1fin}
\end{gather}
This expression is valid at low temperatures and small chemical potential (i.e. smaller than the bandwidth). 
The logarithmic Larmor frequency dependence is not specific to Weyl fermions but is also predicted in a normal metal from the orbital term\cite{knigavko}.
This result agrees with similar calculations in Refs. \onlinecite{maebashi,hirosawa}.

We mention that other part of the hyperfine interaction, which contains both the spin dipole and Fermi contact terms, gives only a 
subleading contribution to the relaxation rate. This can be seen by realizing that
the matrix element of this part of the hyperfine coupling is bounded from above as 
$\|\left({\bf q}\times \left({\bf q}\times \boldsymbol{\sigma}\right)\right)/q^2\|\le 1$, 
and does not diverge for any $\bf q$. Since the wavefunction is also normalized,
this gives a contribution which is smaller than the otherwise leading term. Indeed, using 
Eq. (15) in Ref. \onlinecite{okvatovity}, the spin dipole and Fermi contact terms give
$1/T_1T\sim \max[(k_BT)^4,\mu^4]$ contribution, which, for small $T$ and $\mu$, is negligible with
respect to Eq. \eqref{t1fin}.
Additional pieces of hyperfine coupling, coming from structures on an atomic length scale, also fall into this category and give
similar subleading corrections.

The chemical potential and temperature dependence of $T_1$  in Eqs. \eqref{t1fin} resembles closely to that of graphene\cite{doranmr1}, namely that of 2D Dirac semimetals.
The only difference is the weak Larmor frequency dependence in the Weyl case.
However, these systems are clearly distinguished by their  physical dimensionality, i.e. 3D vs 2D.

Using the archetypical Weyl-semimetal TaP, the nuclear relaxation rate was measured using nuclear quadrupole resonance (NQR) experiments on the Ta nuclear spins\cite{yasuoka}.
The experimental data for $1/T_1T$ exhibits a constant, $T$ independent behaviour at low temperatures, which crosses over to a $T^2$ increase with increasing temperature. This agrees with our analytical results in Eq. \eqref{t1fin}.
However, to account for the fine details of the experimental data, we have to take into account the temperature dependence of the chemical potential.
The experiment was performed at a fixed number of electrons which did not vary with the temperature, which amounts to consider $\mu(T)$ chemical potential.
As we show below, this explains quantitatively all features of the experiment.

\begin{figure}[h!]
\includegraphics[width=8cm]{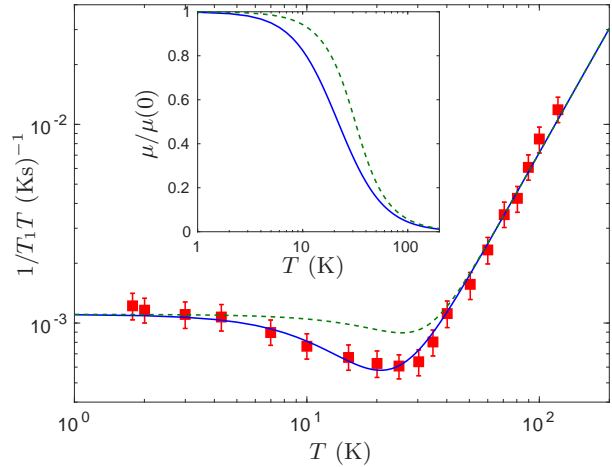}
\caption{The experimental spin-lattice relaxation rate on TaP  from Ref. \onlinecite{yasuoka}
(red squares), together with the theoretical $T_1$ of  Eq. \eqref{weylrelax} using the chemical potential from  Eq. \eqref{exsol} (green dashed line) and also the approximate expression from Eq. 
\eqref{muapprox} (blue line)  with  $\mu(0)/k_B=75$~K and $\hbar\omega_0/k_B=0.0013$~K.
Inset: Temperature dependence of $\mu(T)$ from Eq. \eqref{exsol} (green dashed line) and of the approximate function (blue) with $c=12$.}
\label{figtmu2}
\end{figure}

The total number of electrons in a Weyl semimetal is calculated from the well-known expression\cite{ashcroft}
\begin{equation}
	N(T)=\int d\varepsilon \frac{g(\varepsilon)}{\exp[(\varepsilon-\mu(T))/k_B T]+1},
	\label{nt2}
\end{equation}
where $g(\varepsilon)=\varepsilon^2{V}/{2\pi^2\hbar^3v_F^3}$ is the density of states in Weyl semimetals and $V$ is the volume of the sample.
Using particle number conservation, $N(T\neq 0)-N(T=0)=0$, we get
\begin{equation}
	\int\limits_{-\infty}^{\infty}\varepsilon^2 d\varepsilon\left(\frac{1}{\exp[(\varepsilon-\mu(T))/k_B T]+1}-\Theta(\mu(0)-\varepsilon)\right) =0,
	\label{mut1}
\end{equation} 
where $\Theta(x)$ is the Heaviside function and $\mu(0)$ is the chemical potential at $T=0$. 
Upon evaluating Eq. \eqref{mut1}, we obtain 
\begin{equation}
	\mu^3(T)-\mu^3(0)+\pi^2(k_BT)^2\mu(T)=0.
	\label{mut2}
\end{equation}
This equation has two complex roots, which are irrelevant for our current study, and its real root reads as
\begin{gather}
	\mu(T)=\frac{E(T)}{6}-\frac{2\pi^2(k_BT)^2}{E(T)},
\label{exsol}
\end{gather}
where $E(T)=(108\mu^3(0)+12\sqrt{12 \pi^6(k_BT)^6+81\mu^6(0) })^{1/3}$.
This yields
\begin{gather} 
\frac{\mu(T)}{\mu(0)}\approx\left\{
\begin{array}{cc}
1-\dfrac 13 \left( \dfrac{\pi k_BT}{\mu(0)}\right)^2, & k_BT\ll\mu(0)\\
\left(\dfrac{\mu(0)}{\pi k_BT}\right)^2, & k_BT\gg\mu(0).
\end{array}\right.
\label{mutlimits}
\end{gather}
The $T^2$ initial decrease of the chemical potential is identical to that in a normal Fermi gas\cite{ashcroft} with the 
Fermi energy replacing the chemical potential in the denominator.
In that case, however, the typical Fermi energy scale is $10^4$~K, thus the  $T$ dependence of the chemical potential is 
negligible at the typical energy scales of condensed matter.
On the other hand, for the present case, upon small doping, the temperature dependence of the chemical potential is important and cannot be neglected, since 
as we show below, $\mu(0)$ can be of the order of 10-100~K and even  the $k_BT\gg \mu(0)$ region can easily be reached.

Eq. \eqref{exsol} arises from an ideal Weyl-fermionic band structure, where the linearly dispersing bands extend to arbitrary energies. For any real system, this is clearly not the case as
bands usually terminate at some 
cutoff energy and also display deviations from Eq. \eqref{weylenergy} at higher energies, which requires the explicit knowledge of the full band structure. 
This, in turn, is expected to alter the temperature dependence of the chemical potential.
We model this effect by a phenomenological $\mu(T)$ function, which still preserves the overall features found in the above calculations. To be explicit,  we use 
\begin{equation}
\mu(T)=\frac{\mu(0)}{1+c[k_BT/\mu(0)]^2}.
	\label{muapprox}
\end{equation} 
The experimental data is fitted by plugging Eq. \eqref{muapprox} into Eq. \eqref{weylrelax} using $c$, $\mu(0)$ and the overall scale of $1/T_1T$ as free parameters. 
The experimental data determines roughly $\mu(0)$, which then fixes the scale factor, 
thus the only free fitting parameter is $c$. Other functions than Eq. \eqref{muapprox} with similar asymptotics work equally well.
The result, together with the $\mu(T)$ curve from Eq. \eqref{exsol}
is shown in Fig. \ref{figtmu2}, { giving $c=12$. The phenomenological chemical potential follows closely that of the ideal system from Eq. \eqref{exsol}, 
as shown in the inset of Fig. \ref{figtmu2}.
This encodes all the neglected features of the band structure, including tilting, warping and anisotropy of the Weyl dispersion, as well as deviations from it at high energies.
The scale factor for the relaxation rate is 
$\mu_0\gamma_{n} e/v_F=1.8\times 10^{-14}$~s.
Altogether, a convincing agreement between experiment and theory is reached.} 

{In Ref. \onlinecite{yasuoka}, a phenomenological two-channel relaxation model was used to explain the experimental data. One channel, independent from the Weyl point, 
 was responsible for the initial decrease of $1/T_1T$ with the temperature, 
while the other channel followed an activated Weyl type behaviour as $\sim T^2\exp(-\Delta/k_BT)$, and accounted
for the high $T$ increase of the relaxation rate.
Both the origin and explicit $T$ dependence of the first channel as well as the activation energy $\Delta$ for the Weyl node had been unknown. As opposed to that, our theory
together with the $\mu(T)$ explains all features of the experimental data on the same footing, invoking only the presence of the doped Weyl node.}

{Finally, let us mention that the contribution of the Fermi arcs\cite{fermiarc1,fermiarc2} together with possible topologically trivial surface states is negligible for the
relaxation time. NMR, unlike e.g. ARPES, is a bulk probe and is sensitive to the response of the total volume of the sample. As such, in a typical sample, the surface to volume ratio
is small or in other words, the density of surface states is small compared to the bulk density of states. Therefore, the contribution coming from surface states
is overwhelmed by the bulk contribution.}

\section{Knight shift}

The conduction electrons induce an average static magnetic field through the hyperfine 
interaction at the position of the nucleus, which is associated with the Knight shift \cite{winter,SlichterBook}. As a result,
the nuclear Zeeman energy is given by {$-\hbar\gamma_n BI_z(1+K)$} with $K$ the Knight shift.
A static magnetic field in the $z$ direction cannot depend on the $z$ coordinate, thus its spatial Fourier transform depends only on $q_{x,y}$.
 This follows from that fact that ${\bf B}=[0,0,B(x,y)]$ has 
to satisfy $\nabla{\bf B}=\partial_{z}B_z=0$, so its Fourier transform $B_{\bf q}$ is independent of $q_z$. 

The external magnetic field appears in Eq. \eqref{hamilton} through the vector potential and the Zeeman term. These give rise to an additional perturbation as
\begin{equation}
H'=ev_F \boldsymbol{\sigma}\cdot{\bf A}+\frac{g\mu_B}{2}B_z\sigma_z.
\label{weylhamext}
\end{equation}
Then, the basic question is how this external magnetic field in the vector potential and the Zeeman term in the Hamiltonian of Weyl semimetals influences the nuclear spin through the hyperfine interaction in Eq. \eqref{hfift}.

The effective magnetic field felt by the nuclear spin is obtained by taking the expectation value of Eq. \eqref{hfift} with respect to the electronic degrees of freedom
in the presence of a static magnetic field in the $z$ direction. This gives the energy shift from the orbital part of the hyperfine coupling as
\begin{equation}
\Delta E^o= \mu_0 \gamma_n \hbar e v_F \frac{i I_z }{q^2}\left(q_x \left\langle  \sigma_{y}\right\rangle
-q_y \left\langle\sigma_{x}\right\rangle\right).
\label{knight1}
\end{equation}
Here only the $z$ component of the nuclear spin is relevant since the magnetic field point in the $z$ direction. 
In a similar fashion, the spin part of the hyperfine coupling gives rise to an energy shift as
\begin{equation}
\Delta E^s= \mu_0\gamma_n\hbar\frac{g\mu_B}{2}I_z\langle\sigma_{z}\rangle.
\label{knight2}
\end{equation}
In order  to obtain the Knight shift, we calculate within linear response theory\cite{mahan} in the external magnetic field the quantity 
$\left\langle \sigma_{x,y,z}\right\rangle$ from $H'$ in Eq. \eqref{weylhamext}. {This gives the expectation value of the spin operator in the Weyl
semimetal in the presence of a small magnetic field from $H'$.  
In the absence of this perturbation, all $\left\langle \sigma_{x,y,z}\right\rangle=0$, i.e. the Weyl node 
is not polarized in any direction.}
Since we need the expectation value of the spin operators and both external perturbations, i.e. 
the vector potential $\bf A$ and the Zeeman term $\bf B$ couple to the physical spin of Weyl fermions in Eq. \eqref{weylhamext},
we need the spin-spin correlation function between $\sigma_a$ and $\sigma_b$, denoted as $\Pi^{ab}(\omega=0,{\bf q})$, to determine $\Delta E^o_A$ and $\Delta E^o_B$
from the Kubo formula, respectively.
This is given by
\begin{gather}
\Pi^{ab}({\bf q})=- \frac 1V\sum_{\bf k}\sum_{\lambda,\lambda'=\pm}\frac{f(\varepsilon_{\lambda}({\bf k}))-f(\varepsilon_{\lambda'}({\bf k+q}))}
{\varepsilon_{\lambda}({\bf k})-\varepsilon_{\lambda'}({\bf k+q})}\times\nonumber\\
\times\langle {\bf k},\lambda|\sigma_a|{\bf k+q},\lambda'\rangle\langle {\bf k+q},\lambda'|\sigma_b|{\bf k},\lambda\rangle,
\label{resp6}
\end{gather}
where $f$ is the Fermi function  and the $\omega=0$ limit has already been taken. 
This expression is complex in general due to the complex matrix elements using  Eqs. \eqref{weylfunction}.
{For example, in the case of an external perturbation of the form $\sigma_b\mathcal{F}({\bf q})$, the expectation values are 
$\left\langle \sigma_{a}\right\rangle=-\Pi^{ab}({\bf q})\mathcal{F}({\bf q})$
with $a$, $b$ being $x,y$ or $z$}.

\subsection{Chemical potential dependence at zero temperature}

We expand Eq. \eqref{resp6} in Taylor series in $q$ up to second order. 
After some tedious though straightforward algebra, the spin correlation function is evaluated in this small $\bf q$ limit at $T=0$  as
\begin{gather}
\Pi^{ab} ({\bf q})=\frac{q^aq^b}{12\pi^2\hbar v_F}\left(\ln\left(\frac{W}{|\mu|}\right)-\frac{14}{15}\right)
-\frac{i\epsilon^{abc}q^c\mu}{4\pi^2(\hbar v_F)^2},
\label{totresp}
\end{gather}
where $(a,b,c)$ denotes the spatial direction $(x,y,z)$, $a\neq b$ and $W$ is a sharp high energy cutoff regularizing the theory and $\epsilon^{abc}$ is the Levi-Civita symbol.
We note that while the logarithmic cutoff dependence is expected in the real part of $\Pi^{ab} ({\bf q})$ for any kind of cutoff, i.e. sharp, exponential, gaussian etc.,
the numerical constant, -14/15 is not universal but is expected to be an order one constant for all cutoff schemes.
We also evaluated Eq. \eqref{resp6} numerically and found perfect agreement with Eq. \eqref{totresp}.

Starting with  $\Delta E^o_A$, the Fourier transform of vector potential for a magnetic field in the $z$ direction is represented in different gauges as
\begin{gather}
{\bf A}({\bf q})=\left(0,\frac{B_{\bf q}}{iq_{x}},0\right) \textmd{ or }
{\bf A}({\bf q})=\left(-\frac{B_{\bf q}}{iq_{y}},0,0\right).
\end{gather}
to evaluate $\langle \sigma_{x,y}\rangle$. Since the expectation value $\langle\sigma_{x,y}\rangle$ 
is gauge invariant, it is clear that the vector potential in any gauge can be used
to calculate them, what we use to our favour to simplify the calculations.
 This allows us to write 
{\begin{gather}
\left\langle \sigma_{x}\right\rangle=-e v_F \frac{B_{\bf q}}{iq_x}\Pi^{xy} \textmd{ and }
\left\langle \sigma_{y}\right\rangle=e v_F \frac{B_{\bf q}}{iq_y}\Pi^{yx}
\end{gather}}
using the two distinct gauges.
Substituting it into Eq. \eqref{knight1}, we get
\begin{equation}
\Delta E^o_A=\mu_0 \gamma_n \hbar (e v_F)^2 \frac{I_z B_{\bf q}}{q^2}\left(\frac{q_x}{q_y}\Pi^{yx}+\frac{q_y}{q_x}\Pi^{xy}\right).
\label{respA1}	
\end{equation}
A similar calculation is carried out to consider the effect of the electronic Zeeman term on the spin expectation values, yielding
\begin{equation}
\Delta E^o_B=\mu_0 \gamma_n \hbar e v_F \frac{g\mu_B}{2}
\frac{iI_z B_{\bf q}}{q^2}\left(q_y\Pi^{xz}-q_x\Pi^{yz}\right).
\label{respB1}	
\end{equation}
The spin part of the hyperfine interaction is mostly affected by the magnetic vector potential part of the Weyl Hamiltonian.
This gives
\begin{equation}
\Delta E^s= \mu_0\gamma_n\hbar\frac{g\mu_B}{2}ev_F I_z\Pi^{zx}\frac{B_{\bf q}}{iq_{y}}.
\label{knight3}
\end{equation}
Finally, an additional contribution from the spin part of the hyperfine interaction is in principle possible from the Zeeman term in Eq. \eqref{weylhamext}, involving
the $\chi_{zz}({\bf q})=0$ spin susceptibility. In accord with Ref. \onlinecite{koshino,nomura}, this can in principle
yield a non-universal constant term, independent of both $T$ and $\mu$, which arises entirely from the
high energy part of the spectrum, not taken into account by Eq. \eqref{hamilton}. This constant term can be merged with the chemical shift\cite{abragam}.

Using the spin correlation function in Eq. \eqref{totresp} for Eqs. \eqref{respA1}, \eqref{respB1} and \eqref{knight3} 
and also the fact that $q_z=0$ for a magnetic field in the $z$ direction,
we finally obtain the zero temperature Knight shift as
\begin{gather}
K=\frac{\mu_0  e }{4\pi^2 \hbar }\left(\frac{g\mu_B}{\hbar v_F}\mu-\frac{ev_F}{3}\left[\ln\left(\frac{W}{|\mu|}\right)-\frac{14}{15}\right]\right).
\label{knightmu}
\end{gather}
Here the first term stems from the electronic Zeeman term and is  the paramagnetic contribution, while the second terms arise due to the electronic orbital contribution,
and represents the diamagnetic term.
The logarithmic term, dominating the diamagnetic term, is always negative since $W/ \mu\gg 1$. However,
 the sign of the first, paramagnetic term can change sign depending
on whether the system is electron or hole doped.
 These agree qualitatively with Ref. \onlinecite{koshino}.
This means that already the paramagnetic term can be negative, thus resembling to the diamagnetic contribution, and
by tuning the chemical potential, one can make the Knight shift vanish at some chemical potential or even change its sign.

\subsection{Temperature dependence at $\mu=0$}

The knowledge of the finite temperature spin-spin correlation function in Eq. \eqref{totresp} 
is required to obtain the temperature dependence of the Knight shift. Since it is calculated
from the Kubo formula for non-interacting electrons in Eq. \eqref{resp6}, it depends linearly on the Fermi-Dirac distribution function.
We then use the trick of Ref. \onlinecite{ maldague} for the Fermi function $f(\varepsilon;\mu;T)$ as
\begin{equation}
f(\varepsilon;\mu;T)=\int\limits_{-\infty}^{\infty}d \mu' \left(-\frac{d f(\mu;\mu';T)}{d\mu}\right)\Theta(\mu'-\varepsilon),
\label{fermitrans}
\end{equation}
where $f(\varepsilon;\mu;T)=1/\left(\exp[(\varepsilon-\mu)/k_BT]+1\right)$ and its $T=0$ limit is the Heaviside function as $\Theta(\mu-\varepsilon)$.
Although the expression in Eq. \eqref{resp6} 
is valid for any temperature, only its zero temperature limit is evaluated in Eq. \eqref{totresp}. Nevertheless, using 
the transformation in Eq. \eqref{fermitrans}, the zero temperature response is transformed to finite $T$ by an integral over the chemical potential as
\begin{equation}
\Pi^{ab}(\mu, T)=\int\limits_{-\infty}^{\infty}d \mu' \left(-\frac{d f(\mu;\mu';T)}{d\mu}\right) \Pi^{ab}(\mu', T=0).
\label{resptrans}
\end{equation}

Putting Eq. \eqref{totresp} in Eq. \eqref{resptrans} to get the finite $T$ spin correlator, 
its imaginary part remains unchanged and only its real part
is influenced  by finite temperatures. For $\mu=0$, it reads as
\begin{equation}
Re\Pi^{ab}({\bf q},T)=\frac{q^aq^b}{12\pi^2\hbar v_F}\left(\ln\left(\frac{2e^{\gamma}W}{\pi k_BT}\right)-\frac{14}{15}\right),
\label{realrespt}
\end{equation}
where $\gamma\approx 0.577$ is the Euler-Mascheroni constant.  
Thus, the temperature dependent Knight shift for undoped Weyl semimetals is
\begin{equation}
K=\frac{\mu_0  e }{4\pi^2 \hbar }\left(\frac{g\mu_B}{\hbar v_F}\mu-\frac{ev_F}{3}\left[\ln\left(\frac{2e^{\gamma}W}{\pi k_B T}\right)-\frac{14}{15}\right]\right).
\label{knightt}
\end{equation}

\subsection{Combined effect of temperature and chemical potential}

Combining the finite $T$, $\mu=0$ results  from Eq. \eqref{knightt} with the finite $\mu$, $T=0$ expression in Eq. \eqref{knightmu}, we arrive to our main result. The Knight
shift in Weyl semimetals for any finite doping and temperature scales as
\begin{equation}
K(\mu,T)\approx\frac{\mu_0  e }{4\pi^2 \hbar }\left(\frac{g\mu_B}{\hbar v_F}\mu-\frac{ev_F}{3}\ln\left(\frac{W}{\max[|\mu|,k_B T]}\right)\right),
\label{knightmut}
\end{equation}
and the chemical potential itself is temperature dependent and vanishes gradually with temperature as in Eq. \eqref{mutlimits}.
The first term is interpreted in terms of the Knight shift in normal metals\cite{alloul,abragam}, where $K\sim A_{hf}(\mu) g(\mu)$ with $A_{hf}$ the hyperfine coupling, which is usually energy independent and $g(\mu)$ is the density  of states.
For Weyl semimetals, $g(\mu)\sim \mu^2$, thus an energy dependent hyperfine coupling is required to satisfy this relation as $A_{hf}\sim 1/\mu$.
The effective hyperfine coupling diverges upon approaching the Weyl point and changes sign depending on the doping level. This is in accord with the  analysis of the relaxation time\cite{okvatovity}.

\begin{figure}[!h]
\includegraphics[width=7cm]{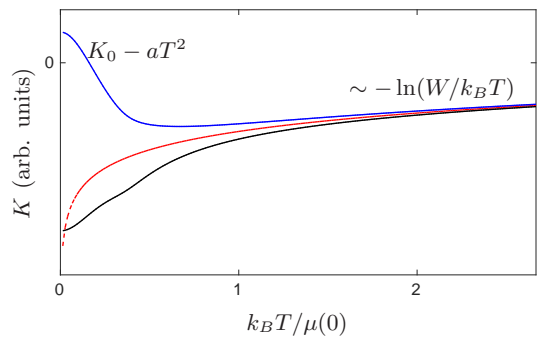}
\caption{Schematic plot of the temperature  dependence of the Knight shift for large positive chemical potential at $T=0$ (blue curve), where the paramagnetic term dominates, for $\mu=0$ (red dashed curve)
and for large negative chemical potential (black curve). While the first case induces a transition from $K>0$ to $K<0$ with increasing temperature, the latter two cases give $K<0$, respectively.}
\label{figknightt}
\end{figure}

Depending on the temperature and the doping level, it can either be dominated by the diamagnetic term with the logarithmic temperature and chemical potential dependence, or by the paramagnetic 
term which can still change sign depending on the electron or hole doping level, respectively.
In typical NMR experiments, the temperature dependence of the relaxation time and the Knight 
shift is measured, because tuning the temperature is an easier task than tuning the chemical potential. 
In Fig. \ref{figknightt}, we show typical behaviours of Knight shift with different zero temperature chemical potentials.

Exactly at the Weyl point, the Knight shift displays strong diamagnetic behaviour and diverges with decreasing temperature as $-\ln(W/k_BT)$.
At $T=0$, Eq. \eqref{knightmu} applies and the sign of the Knight shift is determined by the conspiracy of the paramagnetic and diamagnetic contributions, but for $\mu<0$, it is always negative.
Upon increasing the temperature, two things kick in: first, the chemical potential starts to decrease and the paramagnetic term slowly vanishes as predicted in Eq. \eqref{exsol} and visualized in the inset of Fig. \ref{figtmu2}.
Second, the temperature starts to compete with the chemical potential in the diamagnetic term and for $k_BT>\mu$, it reduces the contribution of the diamagnetic term.
Therefore, at high temperatures $k_BT\gg\mu(0)$, the sign of the Knight shift is most probably negative as the paramagnetic term vanishes due to the vanishing of the chemical potential and only the diamagnetic
contribution remains as  $\sim -\ln(W/k_BT)$. These features are visualized in Fig. \ref{figknightt}.

\section{Korringa relation}

The calculation of the relaxation time $T_1$ and the Knight shift allows to test the validity of the Korringa relation,
i.e. whether $1/T_1TK^2=$const holds. In general, the Korringa relation is valid for a Fermi liquid.
In particular for a non-interacting Fermi-gas\cite{alloul}  
\begin{gather}
\frac{1}{T_1 T K^2}=\frac{4\pi k_B}{\hbar}  \left(\frac{\hbar\gamma_n}{g\mu_B}\right)^2,
\end{gather}
while deviations from this usually indicate certain instabilities, strong correlation effects or transitions.

Since our Weyl fermions are non-interacting, it is interesting to investigate to what extent this Korringa relation holds.
From our results in Eqs. \eqref{t1fin} and \eqref{knightmut}, we infer that while $T_1$ shows rather smooth behaviour and increases roughly with the temperature, the Knight shift exhibits more
intricate behaviour and can even vanish in certain cases, as exemplified in Fig. \ref{figknightt}. This means that $(T_1 T K^2)^{-1}$ can change significantly with both temperature and chemical potential, and can even diverge when
the Knight shift changes sign.

Therefore, it is much more instructive to focus on the $T=0$ behaviour and assume significant doping away from the Weyl point. In this limit, there is a well developed and large Fermi surface,
similar to that in normal metals. In this case, by neglecting the logarithmic terms both in the relaxation time and the Knight shift, we deduce
\begin{gather}
\frac{1}{T_1 T K^2}\approx \frac{4\pi k_B}{3\hbar}  \left(\frac{\hbar\gamma_n}{g\mu_B}\right)^2,
\end{gather}
which is  three times smaller than what is expected in a normal Fermi gas.

Finally, by tuning the system to the close vicinity of the Weyl point with $\mu(T)=0$ or by moving to high temperatures with $k_BT\gg\mu(0)$, it acquires a strong temperature dependence as
\begin{gather}
\frac{1}{T_1 T K^2}\approx\frac{4\pi k_B}{\hbar}  \left(\frac{\hbar\gamma_n}{g\mu_B}\right)^2\times \left(\frac{g\mu_B\pi^{3/2}k_BT}{\hbar e v_F^2}\right)^2
\end{gather}
up to logarithmic corrections in temperature. At the Weyl point, the Korringa relation vanishes for $T\rightarrow 0$ and gets significantly enhanced with the temperature.
Even though the electronic system is non-interacting, the Korringa relation deviates from its ideal value due to the strong temperature
dependence of the spin relaxation time and the very weak temperature dependence of the Knight shift.
The strong spin-orbit coupling, which induces Eq. \eqref{hamilton}, entangles the spin degrees of freedom with the lattice, and spin fluctuations, which play an important
role in determining $T_1TK^2$, causes deviations from the ideal Fermi gas value.

\section{Conclusions}
The purpose of this work is twofold: first, we focused on the spin relaxation time of Weyl fermions in TaP. We took into account the temperature dependence
of the chemical potential, whose characteristic energy scale, separating the high and low temperature behaviour in $\mu(T)$, is the zero temperature
chemical potential, i.e. the Fermi energy of the system, measured from the Weyl point. Unlike in normal metals, this scale can be of the order of 10-100~K for weakly doped Weyl systems, and the temperature
dependence of the chemical potential is essential to understand quantitatively the observed relaxation time.

We also investigated carefully the other characteristics of nuclear magnetic resonance, the Knight shift, which determines the position of the resonance for nuclear spins.
It exhibits rich behaviour as a function of temperature and doping and can even vanish and change sign as a function of these parameters.
Close to absolute zero, it is diamagnetic for small doping,  but can become either positive or negative with increasing doping depending on the doping level (i.e. electron or hole doping).
At high temperature, on the other hand, it is always dominated by the diamagnetic term and decays very slowly as $-\ln(W/k_BT)$ with increasing temperatures.
These unique features, in our opinion, can be used to identify signatures of Weyl points in the band structure even at significantly large doping level.

\appendix

\section{The hyperfine interaction for Bloch wavefunction}

The hyperfine interaction matrix elements should in principle be calculated from Eqs. \eqref{hamilton}
and \eqref{vectorpot} using the full real space Bloch wavefunction.
We show, that its most dominant contribution is captured by using a simple plane wave wavefunction, yielding Eq. \eqref{hfift}.
The real space Bloch wavefunction, corresponding to the upper ($u$) and lower ($l$) spinor component in Eq. \eqref{weylfunction},
is written as\cite{lunde}
\begin{gather}
\Psi_{{\bf k},j}({\bf r})=\exp(i{\bf k\cdot r})u_{{\bf k},j}({\bf r}),
\label{bloch1}
\end{gather}
where $j=u, l$, the $u_{{\bf k},j}({\bf r})=u_{{\bf k},j}({\bf r+R})$ function is lattice periodic with $\bf R$ being the lattice vector.
Using the appropriate atomic wavefunction for Weyl semimetals, it is written as 
$u_{{\bf k},j}({\bf r})=\frac{1}{\sqrt{N}}\sum_{\bf R}\exp(-i{\bf k\cdot (r-R)})\phi_{j}({\bf r-R})$, which is indeed lattice periodic. Here,
$\phi_{j}({\bf r})$ is the atomic wavefunction, $N$ is the total number of lattice sites.
Due to its real space periodicity, it can be Fourier expanded in terms of the reciprocal lattice vectors $\bf G$ as
\begin{gather}
u_{{\bf k},j}({\bf r})=\sum_{\bf G}c_{{\bf k},j}({\bf G})\exp(i{\bf G\cdot r}),
\label{bloch2}
\end{gather}
where $c_{{\bf k},j}({\bf G})$ are the expansion coefficients.
Putting this back to Eq. \eqref{bloch1}, we are in position to discuss the matrix element of any real space operator,  $f_{j',j}({\bf r})$ with $j',j=u$ or $l$, connecting all possible spinor components.
In particular, we are interested in the matrix elements of $f({\bf r})=\bm \sigma \cdot\bf A$ with  $\bf A$ the vector potential from Eq. \eqref{vectorpot}.

By using the Bloch wavefunction from Eq. \eqref{bloch1},
the matrix elements of $f_{j',j}({\bf r})$ are evaluated from
\begin{gather}
\int d{\bf r} \Psi^*_{{\bf k'},j'}({\bf r})
f_{j',j}({\bf r})\Psi_{{\bf k},j}({\bf r})=\nonumber\\
=\int d{\bf r} u^*_{{\bf k'},j'}f_{j',j}({\bf r})\exp(i({\bf k-k')\cdot r})u_{{\bf k},j},
\end{gather}
and upon using Eq. \eqref{bloch2}, this simplifies to
\begin{gather}
\sum_{\bf G,G'}c_{{\bf k},j}({\bf G})c^*_{{\bf k'},j'}({\bf G'})\hat f({\bf q}+\Delta {\bf G}),
\end{gather}
where ${\bf q}=\bf {k-k'}$, $\Delta {\bf G}=\bf {G-G'}$ and $\hat f({\bf q})=\int d{\bf r}f({\bf r})\exp(i{\bf q\cdot r})$.

Applying this to the matrix elements of $\bm \sigma \cdot\bf A$, which are 
essential in determining the orbital part of the hyperfine coupling, they  read as
\begin{gather}
{\bf I}\cdot \sum_{\bf G,G'}c_{{\bf k},j}({\bf G})c^*_{{\bf k'},j'}({\bf G'})\dfrac{\left[\left({\bf q}+\Delta{\bf G}\right)\times\boldsymbol{\sigma}_{j',j}\right]}{|{\bf q}+\Delta{\bf G}|^2},
\label{bloch3}
\end{gather}
where we have dropped unimportant constant prefactors to focus on the ensuing mathematical structure.
Eq. \eqref{bloch3} contains a denominator, which diverges for $\bf G=G'$ and ${\bf q}\rightarrow 0$, but stays finite for $\Delta {\bf G}\neq 0$.
Therefore, we keep only the most dominant, ${\Delta \bf G}=0$ terms from the expansion to focus on the low energy limit of Weyl semimetals in 
Eq. \eqref{hfift}, and neglect the other terms, which are finite in the ${\bf q}=0$ limit.
Moreover, the magnitude of the $\Delta{ \bf G}\neq 0$ terms decreases as $1/|\Delta \bf G|$ with increasing $\Delta \bf G$ from Eq. \eqref{bloch3}, 
therefore
they
give a negligible contribution to the hyperfine coupling.

Since we are interested in the low energy response of Weyl fermions, 
we can also take the $\bf k,k'\ll G$ limit in the expansion coefficients.
The low energy theory, Eq. \eqref{hamilton} is valid in this long wavelength limit.
We can safely take the ${\bf k,k'}=0$ limit in the expansion coefficients, 
because no sharp structures are expected at small wavevector from the Fourier transform of the atomic wavefunction in Eq. \eqref{bloch2}.
Therefore, only an overall, $\bf k$ independent normalization factor as 
$\sum_{\bf G}c_{{\bf 0},j}({\bf G})c^*_{{\bf 0},j'}({\bf G})$ 
remains present for the $\Delta {\bf G}=0$ terms from the Bloch wavefunction, which can be merged with the numerical constants in Eq. \eqref{hfift}.
The resulting expression is then identical to Eq. \eqref{hfift}, which was obtained in Ref. \onlinecite{okvatovity} without the full Bloch wavefunction.

Having identified the most divergent contribution of the hyperfine coupling to Weyl fermions, one can ask whether 
this is the  appropriate one to describe the experimental problem one is dealing with.
The hyperfine coupling, emanating from Eq. \eqref{vectorpot}, is long range in real space therefore it is natural to
consider its long wavelength contribution in momentum space, which is Eq. \eqref{hfift}. Moreover,
as we have shown\cite{okvatovity}, this divergent hyperfine coupling gives the most dominant contribution to the nuclear spin relaxation time and
 is essential to describe the experimentally observed 
$1/T_1T\sim \max[(k_BT)^2,\mu^2]$ type behaviour\cite{yasuoka}. Additional non-divergent (in the ${\bf q}\rightarrow 0$)
hyperfine terms only give rise to subleading, $1/T_1T\sim \max[(k_BT)^4,\mu^4]$ scaling, as  argued below Eq. \eqref{t1fin} as well.
Thus, the additional $\Delta {\bf G}\neq 0$ terms do not explain the experimental data in TaP, only the $\Delta {\bf G}=0$ terms provide the observed behaviour.


\begin{acknowledgments}

This research is supported by the National Research, Development and Innovation Office - NKFIH within the Quantum Technology National Excellence Program (Project No.
      2017-1.2.1-NKP-2017-00001), K108676 and K119442, by the  BME-Nanonotechnology FIKP grant of EMMI (BME FIKP-NAT) and by Romanian UEFISCDI, project number PN-III-P4-ID-PCE-2016-0032.
\end{acknowledgments}

\bibliographystyle{apsrev}

\bibliography{graph_nmr,reference}

\end{document}